\documentstyle[fleqn,aps,prl,twocolumn,graphics,mytitle]{revtex}
\input epsf
\epsfverbosetrue
\begin{document} 

\title{Measurement of Direct Photon Emission in the 
$K_{L}\rightarrow\pi^+\pi^-\gamma$ Decay Mode}

\author{
M.~Arenton$^{13}$,
A.R.~Barker$^{6,**}$,
L.~Bellantoni$^8$,
A.~Bellavance$^{10}$,
E.~Blucher$^5$,
G.J.~Bock$^8$,
E.~Cheu$^1$,
R.~Coleman$^8$,
M.D.~Corcoran$^{10}$,
G.~Corti$^{13}$,
B.~Cox$^{13}$,
A.R.~Erwin$^{14}$,
C.O.~Escobar$^{4}$,
A.~Glazov$^5$,
A.~Golossanov$^{13}$,
R.A.~Gomes$^{4}$,
P.~Gouffon$^{12}$,
K.~Hanagaki$^9$,
Y.B.~Hsiung$^8$,
H.~Huang$^{6}$,
D.A.~Jensen$^8$,
R.~Kessler$^5$,
K.~Kotera$^9$,
A.~Ledovskoy$^{13}$,
P.L.~McBride$^8$,
E.~Monnier$^{5,*}$,
K.S.~Nelson$^{13}$,
H.~Nguyen$^8$,
R.~Niclasen$^6$,
D.G.~Phillips~II$^{13}$,
H.~Ping,$^{14}$,
V.~Prasad$^5$,
X.R.~Qi$^8$,
E.J.~Ramberg$^8$,
R.E.~Ray$^8$,
M. Ronquest$^{13,\dagger}$,
E.~Santos$^{12}$,
J.~Shields$^{13}$,
W.~Slater$^2$,
D.~Smith$^{13}$,
N.~Solomey$^5$,
E.C.~Swallow$^{5,7}$,
P.A.~Toale$^6$,
R.~Tschirhart$^8$,
C.~Velissaris$^{14}$,
Y.W.~Wah$^5$,
J.~Wang$^1$,
H.B.~White$^8$,
J.~Whitmore$^8$,
M.~Wilking$^6$,
B.~Winstein$^5$,
R.~Winston$^5$,
E.T.~Worcester$^5$,
M.~Worcester$^5$,
T.~Yamanaka$^9$,
E.D.~Zimmerman$^6$,
R.F.~Zukanovich$^{12}$\\
(KTeV Collaboration)} 

\address{\vspace{0.1in}
\noindent $^{1}$University of Arizona, Tucson, Arizona 85721 \\
$^{2}$University of California at Los Angeles, Los Angeles, California 90095\\
$^{3}$University of California at San Diego, La Jolla, California 92093\\
$^{4}$Universidade Estadual de Campinas, Campinas, Brazil 13083-970\\
$^{5}$The Enrico Fermi Institute, The University of Chicago, Chicago, Illinois
60637\\
$^{6}$University of Colorado, Boulder, Colorado 80309\\
$^{7}$Elmhurst College, Elmhurst, Illinois 60126\\
$^{8}$Fermi National Accelerator Laboratory, Batavia, Illinois 60510\\
$^{9}$Osaka University, Toyonaka, Osaka 560 Japan\\
$^{10}$Rice University, Houston, Texas 77005\\
$^{11}$Rutgers University, Piscataway, New Jersey 08855\\
$^{12}$Universidade de Sao Paulo, Sao Paulo, Brazil 05315-970\\
$^{13}$The Dept. of Physics and Institute of Nuclear and Particle
Physics, University of Virginia, Charlottesville, Virginia 22901\\
$^{14}$University of Wisconsin, Madison, Wisconsin 53706\\ }

\myabstract{
In this paper the KTeV collaboration reports the analysis of $112.1\times10^3$ candidate
$K_{L}\rightarrow\pi^+\pi^-\gamma$ decays including a background of
671$\pm$41 events with the objective of determining the photon production
mechanisms intrinsic to the decay process. These decays have been analyzed to extract
the relative contributions of the CP violating bremsstrahlung process and the
CP conserving M1 and CP violating E1 direct photon emission processes. The M1
direct photon emission amplitude and its associated vector form factor parameterized as 
$|\tilde{g}_{M1}|(1+ \frac{a_1/a_2}{(M^2_{\rho}-M^2_K)+2M_KE_{\gamma}})$ have
been measured to be $|\tilde{g}_{M1}|=1.198\pm0.035({\rm stat})\pm 0.086({\rm syst})$ and 
$a_1/a_2 = -0.738\pm0.007({\rm stat})\pm 0.018 ({\rm syst})~{\rm GeV}^2/c^2$ respectively. 
An upper limit for the CP violating E1 direct emission amplitude 
$|g_{E1}|\leq 0.21$ (90\%CL) has been found. The overall ratio of direct photon
emission (DE) to total photon emission including the bremsstrahlung process
(IB) has been determined to be DE/(DE +IB) = $0.689\pm 0.021$ for 
$E_{\gamma}\geq$ 20 MeV.\\ \\ 
\vspace{0.25in}
}

\maketitle

The study of the direct photon emission in $K_L\rightarrow\pi^+\pi^-\gamma$
decays gives insight into both the structure of the kaon and the sources 
of CP violation in this mode. This decay proceeds via two main
processes~\cite{ref:6,ref:7}. The
first of these is the inner bremsstrahlung process (IB) in which one of the
charged pions from a CP violating  $K_L\rightarrow\pi^+\pi^-$ decay emits
an E1 electric dipole photon by bremsstrahlung.  The second process is
the emission of a CP violating E1 electric dipole photon or a CP 
conserving M1 magnetic dipole photon together with the $\pi^+\pi^-$ pair
directly from the primary decay vertex.  The photons produced by the
IB process have a typical bremsstrahlung spectrum with $E_{\gamma}$ in the
$K_L$ center of mass peaking toward zero and falling off like $1/E_{\gamma}$,
while the direct photon emission produces an energy spectrum peaked toward
larger $E_{\gamma}$. The M1 amplitude is expected to require a 
form factor $(1+\frac{a_1/a_2}{(M^2_{\rho}-M^2_K)+2M_KE_{\gamma}})$
according to chiral perturbation theory~\cite{ref:chiral} in order to incorporate
the effects of the structure of the $K_L$ on photons emitted at the primary
decay vertex (as opposed to the photons of the bremsstrahlung process emitted
from the charged pions).  In this form factor $M^2_{\rho}$ and $M^2_K$ are
the mass squared of the $\rho$ (770 MeV/c$^2$) and K (497 MeV/c$^2$) mesons..

The KTeV collaboration previously reported a measurement~\cite{ref:1}  using 8669
candidate $K_L\rightarrow\pi^+\pi^-\gamma$ decays accumulated in the
1996 KTeV E832 run at Fermi National Accelerator Laboratory which indicated
clearly the presence of the M1 process and the need for the associated form
factor.  In addition, the presence of M1 photon emission and the need for a
form factor have also been demonstrated by the KTeV E799 experiment~\cite{ref:2,ref:3} and
NA48~\cite{ref:4} measurement of the $K_L\rightarrow\pi^+\pi^-e^+e^-$
mode.  While this mode differs from the $K_L\rightarrow\pi^+\pi^-\gamma$ since the 
photon is virtual converting internally to a $e^+e^-$ Dalitz pair in the
$K_L\rightarrow\pi^+\pi^-e^+e^-$ decay, both modes have the same amplitudes
contributing except for the presence of an extra ``charge radius'' amplitude in the
$K_L\rightarrow\pi^+\pi^-e^+e^-$ decay. Thus we expect the same $g_{M1}$ 
amplitude and associated form factor to be present in the 
$K_L\rightarrow\pi^+\pi^-e^+e^-$ decay. This has been demonstrated by the 
measurements detailed in Refs~\cite{ref:2,ref:3,ref:4}.

Values of $|g_{M1}|$, its form factor, and the ratio
DE/(DE+IB) presented in this paper were determined by using the much larger, 
complete KTeV E832 1997 $K_{L}\rightarrow\pi^+\pi^-\gamma$ data set 
containing $112.1\times 10^3$ candidate $K_L\rightarrow\pi^+\pi^-\gamma$ decays. 
We have analyzed the $K_{L}\rightarrow\pi^+\pi^-\gamma$ decay mode using the
double  differential decay rate
\begin{eqnarray}
\lefteqn{\frac{d\Gamma}{dE_{\gamma}d\cos\theta}=(\frac{E_{\gamma}}{8\pi M_K})^3 (1-\frac{4m_{\pi^2}}{M_{\pi\pi}^2})^{\frac{3}{2}}(1-\frac{2E_{\gamma}}{M_K}) } \\
& & \nonumber \times \sin^2\theta [|E1_{BR}+E1_{direct}|^2 +|M1_{direct}|^2]
\end{eqnarray}
\noindent from the model of Ref.~\cite{ref:5}. In this expression, 
$\theta$ is the angle of the photon with respect to the $\pi^+$ 
in the $\pi^+\pi^-$ center of mass system, $E_{\gamma}$ is the photon energy in 
the $K_L$ rest frame, and 
\begin{eqnarray}
\nonumber\lefteqn{E1_{BR}=(\frac{2M_K}{E_{\gamma}})^2 \frac{|\eta_{+-}|e^{i\Phi_{+-}}e^{i\delta_0}}{1 - (1-\frac{4m^2_{\pi}}{M^2_{\pi\pi}})\cos^2\theta}} \\
\lefteqn{E1_{direct}=|g_{E1}|e^{i\delta_1}} \\
\nonumber\lefteqn{|M1_{direct}|=|\tilde{g}_{M1}|(1+ \frac{a_1/a_2}{(M^2_{\rho}-M^2_K)+2M_KE_{\gamma}})e^{i\delta_1}}
\end{eqnarray}
\noindent where $\delta_0(s=M_K^2)$ and $\delta_1(s=M_{\pi\pi}^2)$ are 
the isospin = 0,1 strong interaction $\pi^+\pi^-$
phase shifts evaluated at the kaon mass and at the particular $\pi^+\pi^-$
mass of a given $K_L\rightarrow\pi^+\pi^-\gamma$ decay. 
$|\eta_{+-}|e^{i\Phi_{+-}}$ is the amplitude for the CP
violating $K_L\rightarrow\pi^+\pi^-$ decay.

Note that there is no interference 
term between the E1 and M1 amplitudes.  However, 
there can still be an interference term in the differential decay rate 
between the $E1_{BR}$ and $E1_{direct}$ amplitudes.  The interference 
will generate a contribution to 
the $E_{\gamma}$ energy spectrum intermediate in energy between the lower energy 
bremsstrahlung photons and the higher energy M1 photons. 

\begin{figure}
 \begin{center}
\epsfxsize=0.80\hsize
\epsfbox{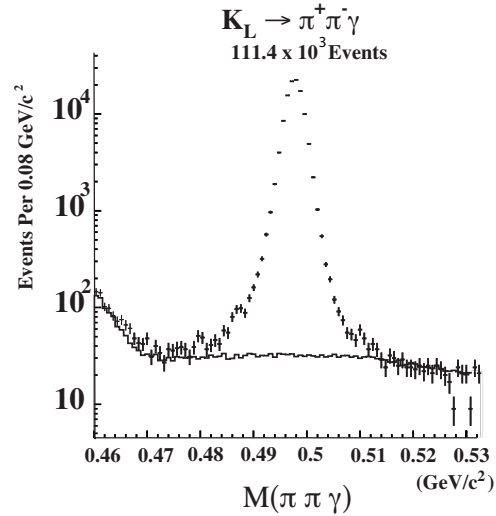}
 \end{center}
\caption{$\pi^+\pi^-\gamma$ invariant mass 
for events passing all $K_{L}\rightarrow\pi^+\pi^-\gamma$ physics cuts except
for the $M_{\pi^+\pi^-\gamma}$ cut.
Crosses are data and the solid line is the fit to the background components}
\label{Fig:2}
\end{figure}

The $K_{L}\rightarrow\pi^+\pi^-\gamma$ signal of $111.4\times 10^3$ events 
above a background of 671$\pm$ 41 events, obtained after the analysis cuts
described below, is shown in Fig.~\ref{Fig:2}. Details of the detector
and beam can be found in Ref.~\cite{ref:9} so we only give a brief
overview here.  A proton beam with a typical intensity of $3\times 10^{12}$ delivered 
in a 20 second spill every minute was incident at an angle of 4.8 mr on a BeO target
producing two nearly parallel $K_{L}$ beams, one of which intercepted a
$K_S$ regenerator and the other of which remained a ``vacuum'' beam.
The data for the $K_{L}\rightarrow\pi^+\pi^-\gamma$ measurement were obtained from
the ``vacuum'' beam decays.  The configuration of 
the KTeV E832 vacuum beam and detector consisted of a vacuum decay 
tube, a magnetic spectrometer with four drift chambers, photon vetoes, a 
Cesium Iodide (CsI) electromagnetic calorimeter, and a muon detector.

Approximately $4.3\times 10^8$ events were extracted from two track 
triggers~\cite{ref:9} by requiring that the two tracks to pass track 
quality cuts and form a vertex with a good vertex $\chi^2$.
These tracks were also required to have opposite charges and ${\rm E/p}\leq~0.85$, 
where E was the energy deposited by the track in the CsI,
and p was the momentum obtained from magnetic deflection. Showers chosen as
photons are required to be far from pion showers and to have a transverse shower shape
consistent with electromagnetic showers.  Only photons
with $E_{\gamma}\geq$ 20 MeV in the $\pi^+\pi^-\gamma$ rest frame were
included in this analysis.

To reduce backgrounds arising from other types of $K_L$ decays in 
which decay products have been missed, the candidate $\pi^+\pi^-\gamma$'s 
were required to have transverse momentum $P_{t}^{2}$ relative to the 
direction of the $K_{L}$ be less than $2.5\times 10^{-4}~{\rm GeV}^2/c^2$ and
$M_{\pi^+\pi^+\gamma}$, the invariant mass of the  $\pi^+\pi^-\gamma$ system,
to be $490~{\rm MeV}/c^2\leq M_{\pi\pi\gamma}\leq 506~{\rm MeV}/c^2$.

The major background to the $K_L\rightarrow\pi^+\pi^-\gamma$ mode was
due to an accidental calorimeter cluster in coincidence with a
$K_L\rightarrow\pi^{\pm}\mu^{\mp}\nu$ decay 
in which the muon was misidentified as a pion. This background was suppressed
by the muon detector identification as well as the $P_{t}^{2}$ and
$M_{\pi\pi\gamma}$ cuts. A smaller background was 
due to $K_L\rightarrow\pi^{\pm}e^{\mp}\nu$ decays 
in which there was an accidental photon, and the electron was misidentified as 
a pion. This background was suppressed by electron E/p identification 
and $P_{t}^{2}$ and $M_{\pi\pi\gamma}$ cuts. The $M_{\pi^+\pi^-\gamma}$ 
spectrum shapes due to the $K_{e3}$ and $K_{\mu 3}$ backgrounds were similar.   

A still smaller background to the $K_L\rightarrow\pi^+\pi^-\gamma$ mode was
$K_{L}\rightarrow\pi^+\pi^-\pi^0$ in which one of the photons from the $\pi^0$ 
decay was not detected in the CsI calorimeter or the photon vetos.  To reduce 
the $K_{L}\rightarrow\pi^+\pi^-\pi^0$ background, the longitudinal momentum
$(P^2_L)_{\pi^0}$ of all candidate $K_L\rightarrow\pi^+\pi^-\gamma$ events 
was calculated (under the assumption that the events were really 
$K_{L}\rightarrow\pi^+\pi^-\pi^0$) in the frame where the $\pi^+\pi^-$
momentum was transverse to the $K_L$ direction. In this frame,$(P^2_L)_{\pi^0}$ is $\geq 0.0$ (except
for resolution effects) for $K_{L}\rightarrow\pi^+\pi^-\pi^0$ decays. In
contrast, the $K_L\rightarrow\pi^+\pi^-\gamma$ decays should have
$(P^2_L)_{\pi^0}\leq~0$.  The requirement $-0.10\leq (P^2_L)_{\pi^0} \leq -0.0055\ {\rm GeV}^2/c^2$,
together with the $P_{t}^{2}$ and $M_{\pi\pi\gamma}$ mass cut, suppressed the
$K_{L}\rightarrow\pi^+\pi^-\pi^0$ background. 

Hyperon decays such as $\Lambda\rightarrow p\pi^-$ plus an
accidental photon with the proton misidentified as a $\pi^+$, or  
$\Xi\rightarrow\Lambda\pi^0$ with a misidentified
proton and one of the $\pi^0$ photons missed, were 
determined to contribute a few events.  Other sources of background 
such as $K_L\rightarrow\pi^+\pi^-$ coincident with
an accidental photon or $K_{S}\rightarrow\pi^+\pi^-\gamma$ produced 
in the neutral beam production target were completely negligible. 

\begin{figure}[h!]
 \begin{center}
\epsfxsize=0.60\hsize
\epsfbox{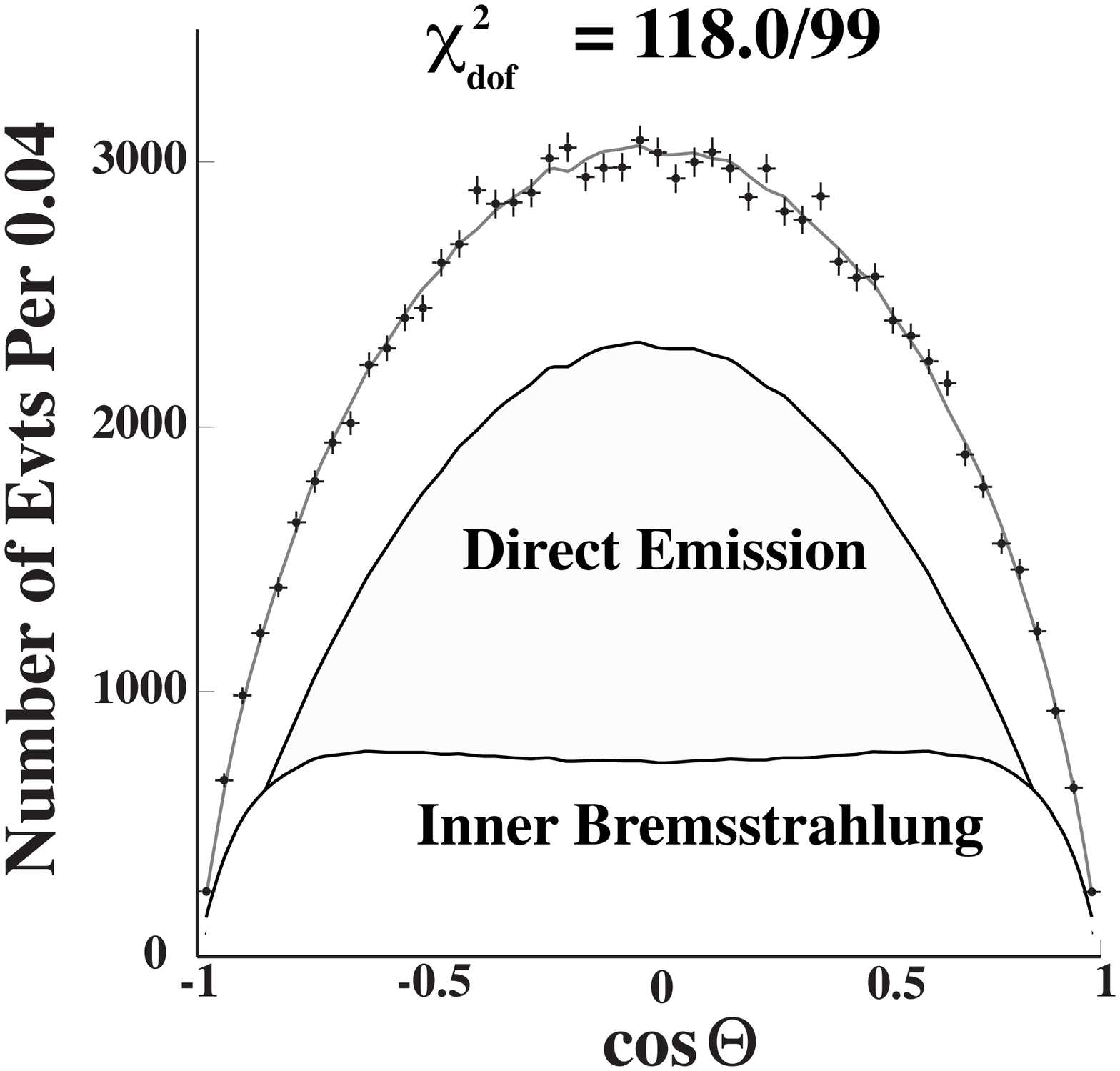}
\epsfxsize=0.60\hsize
\epsfbox{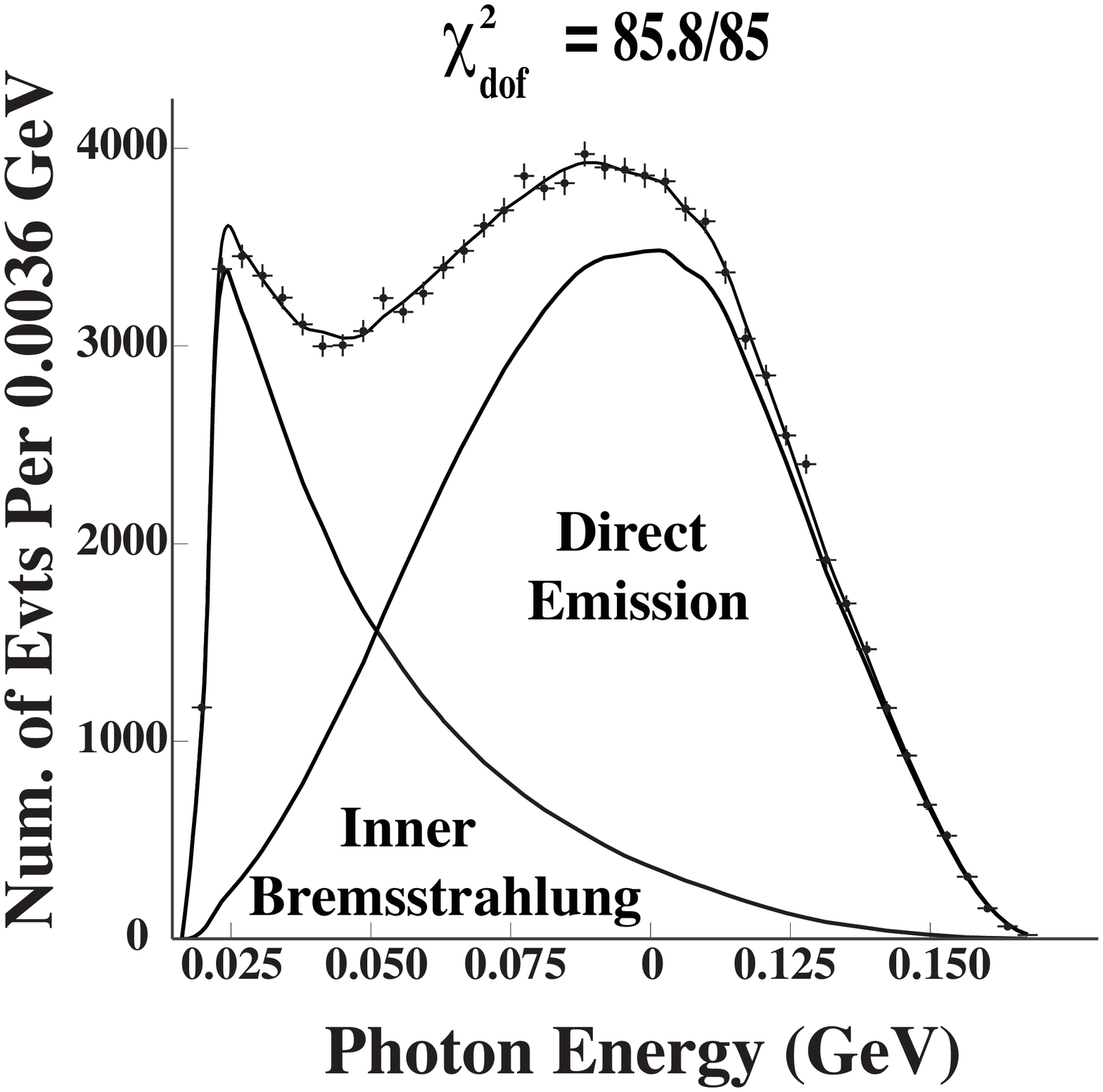}
 \end{center}
\caption{Likelihood fit to the two independent variables in the $K_L$ rest frame:
a) the angle $\theta$ between the $\pi^+$ and the $\gamma$  in the $\pi^+\pi^-$ center of
mass and b) the photon energy spectrum $E_{\gamma}$ in the $K_L$ rest frame.
The components of the photon energy spectrum and the cos$\theta$ spectrum due
to the bremsstrahlung and M1 direct emission processes are shown.}
\label{Fig:3}
\end{figure}

The magnitude of the remnant background after all cuts
was determined by a fit (see Fig.~\ref{Fig:2}) of the sideband  regions
above and below the $K_L$ mass peak to shapes obtained from a Monte Carlo of
the backgrounds leading to an estimated total background of $671\pm41$
events. The best estimate of the composition of this background is 9\%
$K_{L}\rightarrow\pi^+\pi^-\pi^0$, 30\% $K_L\rightarrow\pi^{\pm}e^{\mp}\nu$,
60\% $K_L\rightarrow\pi^{\pm}\mu^{\mp}\nu$, and the remainder due to the other
minor backgrounds mentioned above. The $112.1\times 10^3$ candidate events, 
including the estimated 671 events of background, were analyzed
in a likelihood fit based on equations 1 and 2. The likelihood was a function
of the two independent variables $\theta$ and $E_{\gamma}$, the values of the fit parameters 
$a_1$/$a_2$, $|\tilde{g}_{M1}|$ and $|g_{E1}|$ and nominal values from 
the PDG~\cite{ref:10} for the other model parameters such as $\eta_{+-}$.  
The strong interaction phase shifts of the $\pi^+\pi^-$ system are
taken from Ref.~\cite{ref:11}.  The likelihood was calculated using a 
Monte Carlo event sample generated with nominal values of the
fit parameters, traced through the spectrometer undergoing multiple 
scattering, bremsstrahlung, and secondary decays. The resulting events are
then reconstructed using the same reconstruction
code as was used on data.  These reconstructed Monte Carlo events are then reweighted
with a new set of fit parameters using the $K_L\rightarrow\pi^+\pi^-\gamma$
matrix element of Ref.~\cite{ref:5} and a likelihood is calculate for the new
parameters. The maximum likelihood fits to the two
independent variables cos$\theta$ and $E_{\gamma}$
are shown in Fig.~\ref{Fig:3}a) and Fig.2b) respectively.  


Possible systematic uncertainties in $a_1$/$a_2$, $|\tilde{g}_{M1}|$ and $|g_{E1}|$ due 
to disagreements between data and Monte Carlo  simulations were investigated
by varying analysis cuts and observing variations in these fit parameters. 
In addition, the momentum spectrum of the
$\pi^+\pi^-\gamma$ system observed in the $K_L\rightarrow\pi^+\pi^-$ decays
has been adjusted to agree with the $K_L$ momentum spectrum observed in 
$K^0_L\rightarrow\pi^+\pi^-$ mode and the data has been refit after the
adjustment.  Any differences between $a_1$/$a_2$, $|\tilde{g}_{M1}|$
and $|g_{E1}|$ before and after the final adjustments were taken to be 
a systematic error due to uncertainty in the kaon beam momentum spectrum. 
Systematics due to uncertainties of parameters such as $\eta_{+-}$, and the
strong interaction phase shifts $\delta_{0,1}$ that were not determined by
the fit were studied by varying 
each parameter over $\pm 1\sigma$ of their published values and observing 
the variation of $a_1$/$a_2$, $|\tilde{g}_{M1}|$, and $|g_{E1}|$.
Bremsstrahlung radiation from the pions was studied using the PHOTOS
program~\cite{ref:12}. This radiation could lead to $\pi^+\pi^-\gamma\gamma$
final states in which one of the photons is not observed causing shifts of the
kinematics of the original $\pi\pi\gamma$ decay.
Possible systematic effects due to the non-orthogonality of drift chamber
planes were also studied.  Final overall systematic
errors in $a_1$/$a_2$, $|\tilde{g}_{M1}|$, and $|g_{E1}|$ were obtained by adding 
the individual errors in quadrature.  Table~\ref{systematics} lists the non-zero 
systematic uncertainties of $a_1$/$a_2$ and $|\tilde{g}_{M1}|$, and $|g_{E1}|$.

\begin{figure}[h!]
 \begin{center}
\epsfxsize=1.0\hsize
\epsfbox{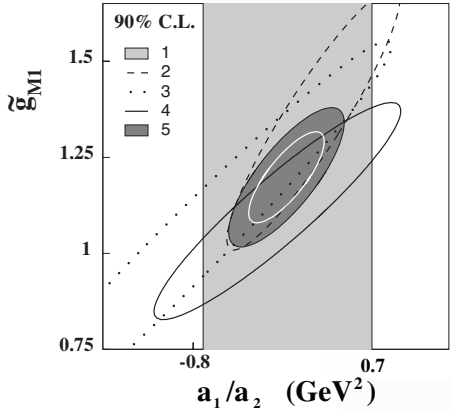}
 \end{center}
\caption{90\% CL contours of \protect{$\tilde{g}_{M1}$} vs. \protect{$a_1/a_2$} for
various experimental measurements; 90\% CL results from
the  \protect{$K_L\rightarrow\pi^+\pi^-\gamma$} mode from this paper (5-shaded
contour) with the 68\% CL contour also shown (white contour); 
For comparison, we show the results from the \protect{$K_L\rightarrow\pi^+\pi^-e^+e^-$} mode 
for NA48 data (3-dotted contour) of Ref.~\protect\cite{ref:4}, 
for KTeV 1997 data (2-dashed contour) in Ref.~\protect\cite{ref:2}, 
and for KTeV 1997+1999 data (4-solid contour) of Ref.~\protect\cite{ref:3}, 
and for \protect{$a_1/a_2$} from earlier KTeV 96
\protect{$K_L\rightarrow\pi^+\pi^-\gamma$} data (1-light gray vertical region) of Ref.~\protect\cite{ref:1}.}
\label{Fig:4}
\end{figure}

The results, including systematic errors of the measurement of the 
M1 direct photon emission amplitude and
the attendant vector form factor, are $a_1/a_2 = (-0.738\pm 0.007(\rm{stat})\pm 0.018 (\rm{syst}))~{\rm GeV}^2/c^2$ and
$|\tilde{g}_{M1}| = 1.198\pm 0.035(\rm{stat})\pm 0.086(\rm{syst})$.  These measurements are in good agreement with
the measurements of Ref.~\cite{ref:1,ref:2,ref:3,ref:4} (see
Fig.~\ref{Fig:4}). After incorporating the systematic errors, an upper limit of
$|g_{E1}| \leq 0.21$ (90\% CL) was obtained. 

Using the result for $|\tilde{g}_{M1}|$ and its associated form factor and 
taking $|g_{E1}|$ to be equal to zero, the ratio of direct to total
photon emission in $K_L\rightarrow\pi^{+}\pi^{-}\gamma$ decay was determined
by integrating the M1 and bremsstrahlung processes over $\theta$ and
$E_{\gamma}$ (for $E_{\gamma}\geq$ 20 MeV) to be DE/(DE+IB)= $0.689\pm0.021$.
This result is in good agreement with Ref.~\cite{ref:1}.


In conclusion, this paper presents the best measurements achieved to
date for the M1 direct photon emission form factor parameters 
$|\tilde{g}_{M1}|=1.198\pm0.035({\rm stat})\pm 0.086({\rm syst})$ 
and $a_1/a_2 -0.738\pm0.007({\rm stat})\pm 0.018 ({\rm syst})~{\rm GeV}^2/c^2$
in the $K_{L}\rightarrow\pi^+\pi^-\gamma$ 
and $K_{L}\rightarrow\pi^{+}\pi^{-}e^{+}e^{-}$ decay modes.  These measurements
are in good agreement with our previous measurement of $a_1/a_2$ using the 1996 KTeV
$K_L\rightarrow\pi^{+}\pi^{-}\gamma$ data~\cite{ref:1} and with our
measurements of $|\tilde{g}_{M1}|$ and $a_1/a_2$ using the 1997 and 1999 
KTeV $K_{L}\rightarrow\pi^{+}\pi^{-}e^{+}e^{-}$ data~\cite{ref:2,ref:3} and with
NA48 results~\cite{ref:4} from
$K_{L}\rightarrow\pi^{+}\pi^{-}e^{+}e^{-}$. We have also determined an upper 
limit $|g_{E1}|\leq 0.21$ (90\%CL) for CP violating E1 direct photon emission in the 
$K_L\rightarrow\pi^{+}\pi^{-}\gamma$ mode consistent with that measured 
using $K_L\rightarrow\pi^{+}\pi^{-}e^+e^-$ decays~\cite{ref:3}.  

We thank the FNAL staff for their contributions.  This work was supported 
by the U.S. Department of Energy, the U.S. National Science Foundation, 
the Ministry of Education and Science of Japan, the Fundao de Ampaaro
Pesquisa do  Estado de So Paulo-FAPESP, the Conselho Nacional de
Desenvolvimento Cientifico e Technologico-CNPq, and the CAPES-Ministerio da Educao. 

\vspace{0.25cm}
\noindent $^{\dagger}$ To whom correspondence should be addressed.\\
\noindent Electronic address: ronquest@uvahep.phys.virginia.edu \\
\noindent $^{*}$ Permanent address C.P.P. Marseille/C.N.R.S., France\\
\noindent $^{**}$ Deceased\\

\begin{table}
\begin{tabular}{|l|c|c|c|c|}
Source                             & $|\tilde{g}_{M1}|$  & $a_1/a_2$ &  $|g_{E1}|$    \\
\hline
Differing initial MC parameters        & 0.0093            &  0.0021     &   0.013 \\
Kaon Beam Momentum Uncertainty         & 0.0031            &  0.0004     &   0.005 \\
Background uncertainty                 & 0.0355            &  0.0067     &   0.045  \\
Pion bremsstrahlung                    & 0.0326            &  0.0140     &   0.097  \\
Non-Orthogonality of chambers          & 0.0402            &  0.0013     &   0.009 \\
Physics cut variations                 & 0.0463            &  0.0056     &   - \\
Fitting resolution                     & 0.014             &  0.0056     &   0.024 \\
$E_{\gamma}$,cos$\theta$ resolution    & 0.023             &  0.0042     &   0.038 \\
$\eta_{+-}$ uncertainty                & 0.0171            &  0.0014     &   - \\
$\delta_0$ phase uncertainty           & 0.0111            &  0.0021     &   - \\
$\delta_1$ phase uncertainty           & 0.0053            &    -        &   - \\
\hline
Total Systematic Error                 & 0.086             &  0.018     & 0.117   \\
\end{tabular}
   \caption{Contributions to the systematic errors (67\% CL) for $a_1$/$a_2$, $|\tilde{g}_{M1}|$, and
$|g_{E1}|$.} 
 \label{systematics}
\end{table}


\begin{thebibliography}{99}
\bibitem{ref:6} L.M. Sehgal and M. Wanninger, Phys. Rev. \textbf{D46},
1035(1992); {\em ibid.} \textbf{D46}, 5209(E)(1992).
\bibitem{ref:7} P. Heiliger and L.M. Sehgal, Phys. Rev. \textbf{D48},
4146(1993).
\bibitem{ref:1} A. Avati-Harati {\em et al.},Phys. Rev. Lett.  \textbf{86},
761(2001).
\bibitem{ref:chiral} Y.C.R. Lin and G. Valencia, Phys. Rev. \textbf{D37}, 143(1988).
\bibitem{ref:2} A. Avati-Harati {\em et al.}, Phys. Rev. Lett. \textbf{84},
408(2000).
\bibitem{ref:3} E. Abouzaid {\em et al.}, hep-ex/05080010.
\bibitem{ref:4} A. Lai  {\em et al.} Eur. Phys. Jou. \textbf{C30}, 33(2003).
\bibitem{ref:5} L.M. Sehgal and J. van Leusen, Phys. Rev. Lett. \textbf{83},
4933(1999).

\bibitem{ref:8} J.K. Elwood, M.B. Wise, and M.J. Savage, Phys. Rev. \textbf{D52},
5095(1995);  J.K. Elwood {\em et al., ibid.}, \textbf{D53},
2855(E)(1996); J.K. Elwood {\em et al., ibid.}, \textbf{D53},
4078(1996).
\bibitem{ref:9} A. Avati-Harati {\em et al.}, Phys. Rev. \textbf{D67},
012005(2003); {\em ibid.} \textbf{D70}, 079904(2004).
\bibitem{ref:10} S. Eidelman {\em et al.}, Phys. Lett. \textbf{B592},
010001(2004).
\bibitem{ref:11} S. Pislak {\em et al.}, Phys. Rev. Lett. \textbf{87},
221801(2001).
\bibitem{ref:12} E. Barberio and Z. Was, Comput. Phys. Commun. \textbf{79}, 291(1994).
\end{thebibliography}
\end{document}